# Electron nematic fluid in a strained $Sr_3Ru_2O_7$ film

Patrick B. Marshall[a], Kaveh Ahadi, Honggyu Kim, and Susanne Stemmer[b]

Materials Department, University of California, Santa Barbara, CA 93106-5050, USA

[a] Electronic mail: pmarshall@mrl.ucsb.edu

[b] Electronic mail: stemmer@mrl.ucsb.edu

**Abstract**

Sr$_3$Ru$_2$O$_7$ belongs to the family of layered strontium ruthenates and exhibits a range of unusual emergent properties, such as electron nematic behavior and metamagnetism. Here, we show that epitaxial film strain significantly modifies these phenomena. In particular, we observe enhanced magnetic interactions and an electron nematic phase that extends to much higher temperatures and over a larger magnetic field range than in bulk single crystals. Furthermore, the films show an unusual anisotropic non-Fermi liquid behavior that is controlled by the direction of the applied magnetic field. At high magnetic fields the metamagnetic transition to a ferromagnetic phase recovers isotropic Fermi-liquid behavior. The results support the interpretation that these phenomena are linked to the special features of the Fermi surface, which can be tuned by both film strain and an applied magnetic field.

# I. Introduction

The family of quasi-two-dimensional strontium ruthenate compounds ($Sr_{n+1}Ru_nO_{3n+1}$, where $n$ is the number of perovskite cells alternating with SrO layers in the Ruddlesden-Popper structure) hosts an exceptionally rich array of unconventional electronic and magnetic phenomena [1-3]. The $n = 2$ member, $Sr_3Ru_2O_7$, exhibits magnetic and electronic instabilities that are connected to a magnetic-field-tuned criticality of the Fermi surface [4]. $Sr_3Ru_2O_7$ is a Fermi liquid that is in close proximity to various magnetic instabilities [5]. In particular, neutron scattering reveals fluctuations of a ferromagnetic order parameter below 150 K that transition to antiferromagnetic fluctuations below 20 K [6]. Under magnetic fields (~ 5 T), a metamagnetic transition from a low-magnetization state to a ferromagnetic state is observed [4]. Around the metamagnetic transition a maximum in the magnetoresistance is observed under in-plane magnetic fields (parallel to the planes containing the $RuO_6$ octahedra). Furthermore, non-Fermi liquid behavior and a quantum critical "fan" appear in the phase diagram as the magnetic field is varied [2,4]. Below 1 K, an electron nematic phase [1], in which the resistance depends on the direction and magnitude of an in-plane magnetic field, has generated significant interest [7]. Recently, it has been shown that it is likely associated with a field-controlled spin density wave (SDW) [8].

What makes the layered strontium ruthenates especially interesting is that many phenomena can be closely linked to special shapes of their Fermi surfaces, such as nesting [9] and singularities in the density of states near the Fermi level [10-12]. The ability to precisely control electronic and magnetic interactions by subtle modifications of the Fermi surface makes $Sr_3Ru_2O_7$ an interesting platform to elucidate the pathways to novel electronically or magnetically ordered phases, quantum criticality, and their relationship to the underlying band structure. In particular, subtle modifications of the Fermi surface by epitaxial thin film strains should influence the dynamics of these magnetic field controlled instabilities and potentially give rise to new phases.

In this study, we report on transport properties and magnetization studies of compressively-strained epitaxial $Sr_3Ru_2O_7$ thin films grown using a recently developed hybrid molecular beam epitaxy (MBE) technique [13]. Low-temperature magnetotransport measurements reveal anisotropies that depend on the relative orientations of magnetic field and current, consistent with the emergence of spin-density wave order that persists to higher temperatures than those reported for single crystals. The itinerant antiferromagnetism also gives rise to anisotropies in the temperature-dependent scattering rates of the in-plane resistivity, resembling a non-Fermi liquid state where critical fluctuations only affect certain regions of the Fermi surface with a momentum that is controlled by the applied field.

## II. Experimental

$Sr_3Ru_2O_7$ thin films were grown by MBE on (001) $(LaAlO_3)_{0.3}(Sr_2AlTaO_6)_{0.7}$ (LSAT) substrates using a hybrid MBE technique described elsewhere [13]. The flux of Sr was reduced relative to the flux used for the growth of $Sr_2RuO_4$ in Ref. [13], while the flux of the Ru-containing gas source was held constant. This favors the growth of the $Sr_3Ru_2O_7$ phase, in which the ratio of Sr to Ru is lower. Growth was carried out in an oxygen plasma with a background pressure of $3\times10^{-6}$ Torr and a substrate thermocouple temperature of 950 °C. The thickness of the $Sr_3Ru_2O_7$ film from which the measurements were taken in this study was 20 nm. Contacts consisting of 400 nm of Au on 40 nm of Ti were deposited in a van der Pauw geometry using an electron-beam evaporator. In-plane sheet resistance and magnetotransport measurements were performed in a Quantum Design Physical Property Measurement system. The magnetotransport measurements were made with the applied magnetic field (*B*) in the *ab*-plane of the film along the <100> family of directions. Measurements of the resistance were made along both the [100] and [010] directions, providing measurements both parallel and transverse to the applied field. Magnetization measurements were carried out in a Quantum Design SQUID magnetometer.

## III. Results

X-ray diffraction (Fig. 1) indicated phase pure films. The 002, 004, 00$\underline{10}$, and 00$\underline{12}$, and 00$\underline{14}$ Bragg peaks are present. The low structure factors of the 006 and 008 reflections and low film thickness prevent their peaks from being resolved. A cross-section high-resolution transmission electron microscopy image is shown in Fig. 2 and confirmed the absence of misfit dislocations at the interface, indicating that the films were coherently strained to the LSAT. The only defects detected were occasional regions of the $n = 3$ structure intertwined in $n = 2$ phase. Growth took place along the [001] direction, with the *c*-axis out of plane.

$Sr_3Ru_2O_7$ films are metallic down to 2 K [Fig. 3(a)]. The temperature derivative of the resistivity ($\rho$), d$\rho$/d*T*, as a function of temperature is shown in Fig. 3(b). Two abrupt changes in d$\rho$/d*T*, near 165 K and 40 K, respectively, can be seen. They coincide with changes in the magnetic behavior. Figure 4 shows the magnetization as a function of magnetic field, measured at 50 K and 2 K, respectively. At 50 K, a small hysteresis is present with a coercive field of 100 Oe, indicating weak ferromagnetic order. As the temperature is reduced further, the coercivity vanishes, suggesting that the magnetism has transitioned to antiferromagnetic order.

Figure 5 shows the magnetoresistance at several temperatures for configurations with the current (*I*) in the direction of the in-plane applied magnetic field, $B \parallel I$ (parallel configuration), and with the current perpendicular to the field, $B \perp I$ (transverse configuration). At 110 K negative, quasi-linear magnetoresistance is observed in both configurations, signifying the ferromagnetic phase. Below 15 K, strong anisotropy begins to set in, with the low-field magnetoresistance remaining negative in the parallel configuration but becoming positive in the transverse configuration. As the temperature is lowered further, prominent peaks around 5 T arise in the transverse configuration, while they are greatly suppressed in the parallel configuration. Beyond 5 T the behaviors of the two configurations are the same, with a sharp quasi-linear negative magnetoresistance marking the completion of the metamagnetic transition and the onset of long-range ferromagnetic order.

The temperature dependence of the normalized resistances, $R(T)/R(2\ K)$, is shown in Figs. 6 (a-c) with applied in-plane magnetic fields of 0 T, 5 T, and 9 T. Results from both parallel and transverse configurations are shown. In the absence of an applied magnetic field, similar characteristics are observed, with the curves nearly falling on top of one another. At the critical magnetic field of 5 T enhanced scattering is observed in the transverse configuration, while the behavior in the parallel configuration is unchanged. At 9 T the scattering in the transverse configuration is suppressed and isotropic behavior is recovered. Figures 6 (d-f) show $\log(dR/dT)$ vs. $\log(T)$ for the same data. Assuming a power law dependence of the resistivity of the form $R = R_0 + AT^\alpha$, the power law exponent $\alpha$ can be extracted using the slope of these plots as $\alpha = slope + 1$. The black dotted lines are fits with $\alpha$ set to 2 to demonstrate where Fermi liquid behavior is obeyed. The exponent is Fermi liquid-like ($\alpha \approx 2$) for both configurations if there is no applied field. At 5 T, the parallel configuration retains Fermi-liquid behavior while the transverse configuration displays $\alpha \approx 1.6$ at the lowest temperatures. Additionally, the intercept of the transverse configuration, $\log(\alpha A)$, is markedly increased at 5 T. This indicates an enhancement of the effective mass, as $A \propto m^*$. At high magnetic fields (> 9 T), isotropy is recovered, with the resistance in the transverse configuration returning to Fermi liquid behavior and a decreased effective mass.

## IV. Discussion

The weak ferromagnetism observed in the intermediate temperature range (20 K - 150 K) contrasts with bulk $Sr_3Ru_2O_7$, which shows only short-range ferromagnetic fluctuations at these temperatures. Below 20 K the magnetization hysteresis vanishes and the magnetotransport properties become strongly dependent on the magnitude and direction of the external field, indicating an electron nematic phase. Spin density waves (SDWs) are believed to be the origin of electronic nematicity in $Sr_3Ru_2O_7$ [8]. SDWs cause anisotropies in the resistance because they remove of states from portions of the Fermi surface as determined by the SDW wave vector. Anisotropy in the magnetoresistance with respect to parallel and perpendicular fields, as observed here, arises if the external magnetic field controls the orientation of the SDW. In bulk $Sr_3Ru_2O_7$, two SDW phases form narrow domes in *B-T* space with a maximum critical temperature of only 1 K [8]. In contrast, the SDW and nematicity in the thin film sample are already present at low fields and at an order of magnitude higher temperature (20 K), where bulk $Sr_3Ru_2O_7$ only shows antiferromagnetic fluctuations [6]. To illustrate this point, Fig. 7 shows a phase diagram, mapping out the region of *B-T* space in which nematic order is present in the strained thin film sample. The experimental data points shown in Fig. 7 correspond to the position of the maxima in the magnetoresistance at each temperature. In addition to the extended range of temperatures and magnetic fields, the nature of the SDW order is likely different, as an in-plane magnetic field component caused an increase in the resistance in the parallel direction in the bulk samples. The peak in the magnetoresistance around 5 T signifies the metamagnetic transition, which occurs at a similar field as in bulk $Sr_3Ru_2O_7$. While non-Fermi liquid behavior around the metamagnetic transition has been observed in bulk $Sr_3Ru_2O_7$ [2,4], here it is found to be strongly anisotropic *and* controlled by the magnetic field direction. This distinguishes $Sr_3Ru_2O_7$ from other materials with anisotropic breakdown of the Fermi liquid behavior that is tied to specific crystallographic directions (for examples, see refs. [14,15]). Above 5 T isotropic, quasi-linear negative magnetoresistance takes over, which is strikingly similar to that in the intermediate-temperature, weakly ferromagnetic phase, and the films show isotropic Fermi liquid behavior.

Epitaxial film strain is the most likely origin for the observed differences between bulk and thin films. Even a modest amount of uniaxial pressure (0.1 GPa) can induce ferromagnetic order in bulk $Sr_3Ru_2O_7$ [16]. The epitaxial film in this study was coherently strained to the LSAT substrate ($a$ = 3.868 Å), yielding a biaxial compressive stress of 1.1 GPa, as calculated from the elastic tensor of $Sr_3Ru_2O_7$ [17] and assuming a pseudocubic lattice parameter of $Sr_3Ru_2O_7$ of 3.888 Å). Epitaxial film strain affects the shape of the Fermi surface and the proximity of the Fermi level to singularities, as has, for example, been shown for the $n$ = 1 member of the series, $Sr_2RuO_4$ [18]. SDWs are typically associated with nested Fermi surfaces. The appearance of a nematic (SDW) phase that exists over a much larger field range and at much higher temperatures than SDWs reported for bulk $Sr_3Ru_2O_7$ indicates that the compressive epitaxial strain promotes Fermi surface nesting. The metamagnetic transition in bulk $Sr_3Ru_2O_7$ has been shown to be due to a van Hove singularity near the Fermi level [10,12,19]. In a simple picture, Zeeman splitting under a magnetic field can transfer additional spectral weight to the Fermi level, when it is located near a singularity, thereby exceeding the limit for the Stoner criterion for ferromagnetism. Theoretical calculations show that the critical temperature and field for metamagnetism are directly correlated with the proximity of the Fermi level to the van Hove singularity [10].

Furthermore, metamagnetic transitions between nematic and isotropic phases are correlated with a rapid change in the Fermi surface topology (Lifshitz transition) [20,21]. The fact that the anisotropic non-Fermi liquid behavior, enhanced scattering (magnetoresistance peak), and mass enhancement near the metamagnetic transition all depend on the orientation of the magnetic field confirms the strong connection between the emergent properties of $Sr_3Ru_2O_7$ and field-controlled instabilities of the Fermi surface, which are modified by epitaxial strain.

Finally, we comment on the quasi-linear negative magnetoresistance that is observed in the ferromagnetic regimes. The fact that it is very similar in two different parts of the temperature-magnetic field space suggests that it is intrinsic and related to the Fermi surface in the isotropic, ferromagnetic regime. Some theoretical models have suggested a transition between open and closed Fermi surfaces at the metamagnetic transition [21], while others have suggested that negative magnetoresistance can arise when the majority spin gets shifted from a flat into a more dispersive part of the Fermi surface at the magnetic transition [22]. These scenarios would be consistent with the data, and further establish a strong correlation between the magnetotransport properties and the Fermi surface in $Sr_3Ru_2O_7$.

## V.   Conclusions

To summarize, it was shown that epitaxial $Sr_3Ru_2O_7$ films exhibit qualitatively similar types of emergent phenomena as have been observed in bulk crystals, including a nematic phase and non-Fermi liquid behavior. Epitaxial thin film strain significantly modifies these phenomena and expands the temperature and magnetic field range in which they are observed. It also gives rise to behavior not observed in bulk crystals, such as an anisotropic non-Fermi liquid that is controlled by the magnetic field direction. The study highlights the extraordinary sensitivity of these phenomena to details for the Fermi surface, opening up the possibility of using strain engineering to gain an improved understanding of a wide range of emergent phenomena.

## Acknowledgements

The work was supported by a MURI funded by the U.S. Army Research Office (Grant No. W911NF-16-1-0361). The work made use of the MRL Shared Experimental Facilities, which are supported by the MRSEC Program of the U.S. National Science Foundation under Award No. DMR 1720256.

**Figure Captions**

**Figure 1:** High-resolution x-ray diffraction scan of a $Sr_3Ru_2O_7$ film grown epitaxially on (001) LSAT. The substrate peaks, marked by orange squares, correspond to the 001, 002, and 003 Bragg reflections of the LSAT substrate from left to right, respectively. The film peaks, marked by red circles, correspond to the 002, 004, 00$\underline{10}$, 00$\underline{12}$, and 00$\underline{14}$ Bragg peaks of $Sr_3Ru_2O_7$. The absence of the 006 and 008 peaks can be explained their low structure factors. The film thickness of 20 nm prevents them from being observed. The absence of any additional peaks demonstrates phase purity.

**Figure 2:** (a) High-resolution transmission electron microscopy image of a 20 nm thick $Sr_3Ru_2O_7$ film grown epitaxially on an (001) LSAT. The green line represents the growth orientation, which is along the [001] crystallographic direction (c-axis out of plane). The absence of misfit dislocations demonstrates that the film is coherently strained to the substrate. The region in the green box (b) is an enlarged image showing the bilayer structure of the $n = 2$ phase $Sr_3Ru_2O_7$. The region in the yellow box (c) shows a defect region with $n = 3$, which were observed occasionally throughout the film.

**Figure 3:** (a) In-plane resistivity ($\rho$) as a function of temperature. Metallic behavior is observed at all temperatures down to 2 K. (b) Derivative, $d\rho/dT$, as a function of temperature. Features arising from magnetic transitions are seen at 165 K and 40 K, respectively.

**Figure 4:** Magnetization ($M$) as a function of magnetic field ($B$) at (a) 50 K and (b) 2 K. The coercive field of 100 Oe at 50 K is indicative of weak long-range ferromagnetic order, while the absence of any hysteresis at 2 K suggests that the magnetism has transitioned to a antiferromagnetic order.

**Figure 5:** Normalized (to $B = 0$) magnetoresistance at several temperatures for both (a) $B \perp I$ and (b) $B \parallel I$. At temperature greater than 110 K isotropic, negative magnetoresistance signifies the ferromagnetic phase, while anisotropy sets in below 15 K. At low temperatures the positive magnetoresistance peak associated with the metamagnetic transition is prominent with $B \perp I$, while it is greatly suppressed with $B \parallel I$.

**Figure 6:** (a-c) Normalized in-plane resistivity, $R(T)/R(2 K)$, for $B \perp I$ and $B \parallel I$ under applied fields of 0 T, 5 T, and 9 T. The temperature dependence is nearly isotropic in the absence of an applied field, while enhanced scattering is observed at 5 T with $B \perp I$. Isotropy is recovered at 9 T. (d-f) Log $(dR/dT)$ vs. log $(T)$ plotted for the same data as in (a-c). The black dotted lines in are fits with $\alpha$ set to 2, to more clearly show where Fermi liquid behavior is obeyed.

**Figure 7:** Phase diagram showing the $B$-$T$ phase space in which nematic behavior is observed in the strained $Sr_3Ru_2O_7$ film. The points correspond to the positions of the maxima in the magnetoresistance at each temperature. The range of magnetic fields and temperatures over which nematic behavior is observed is significantly extended relative to previous studies on bulk samples.

**Figure 1**

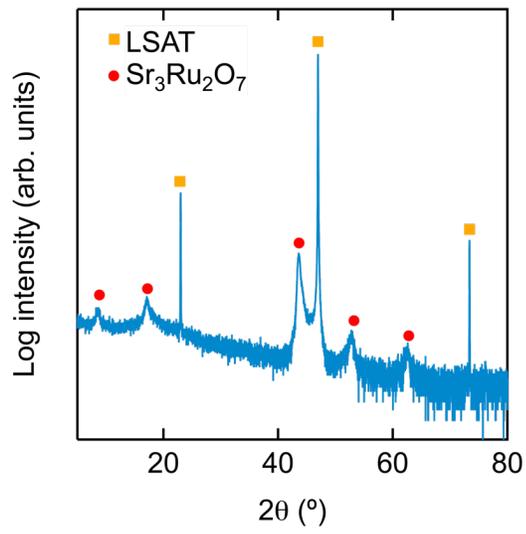

**Figure 2**

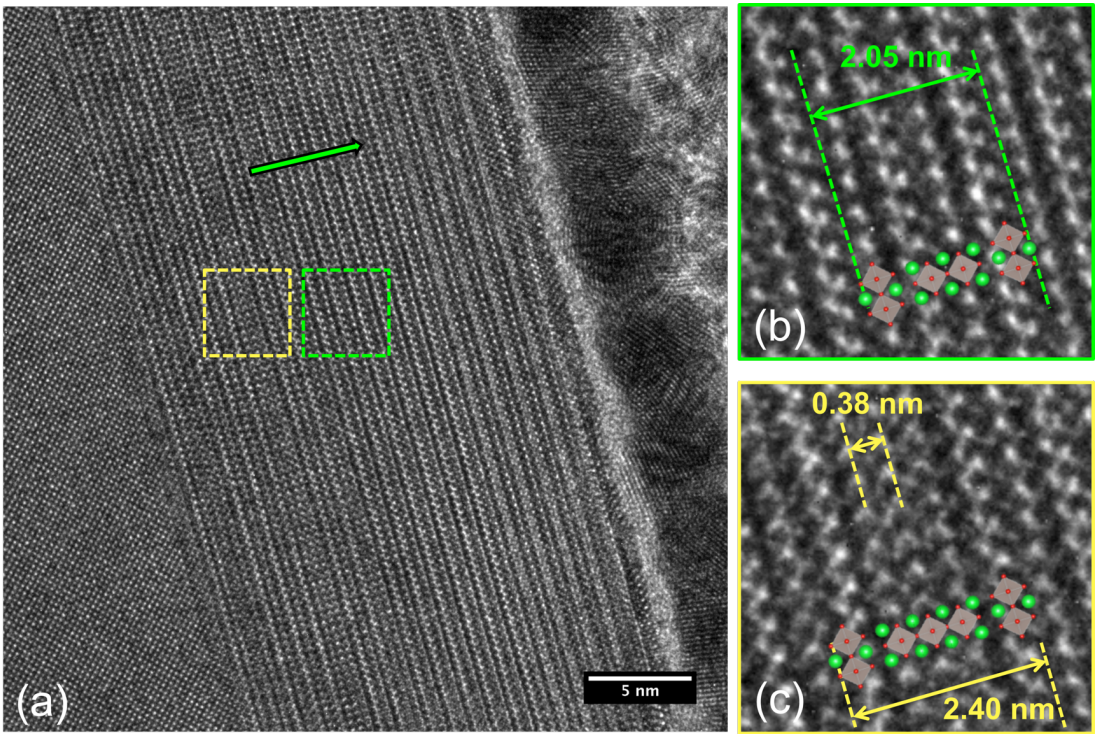

**Figure 3**

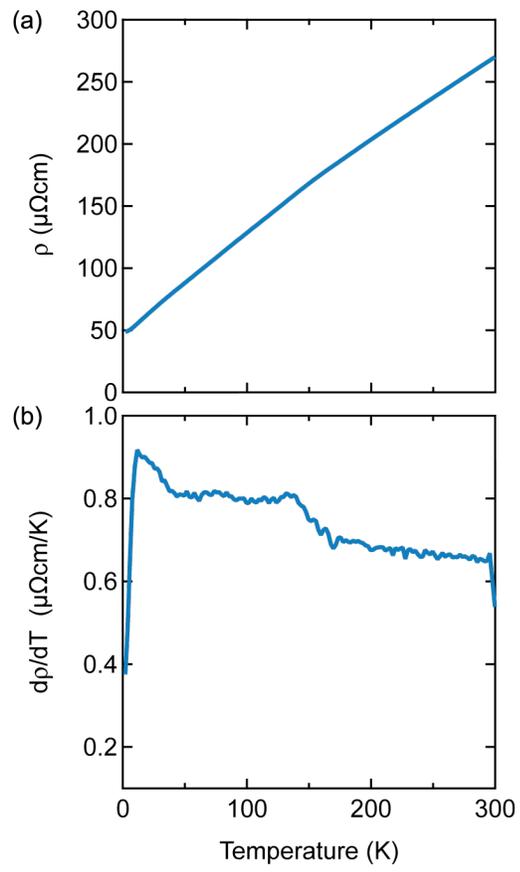

**Figure 4**

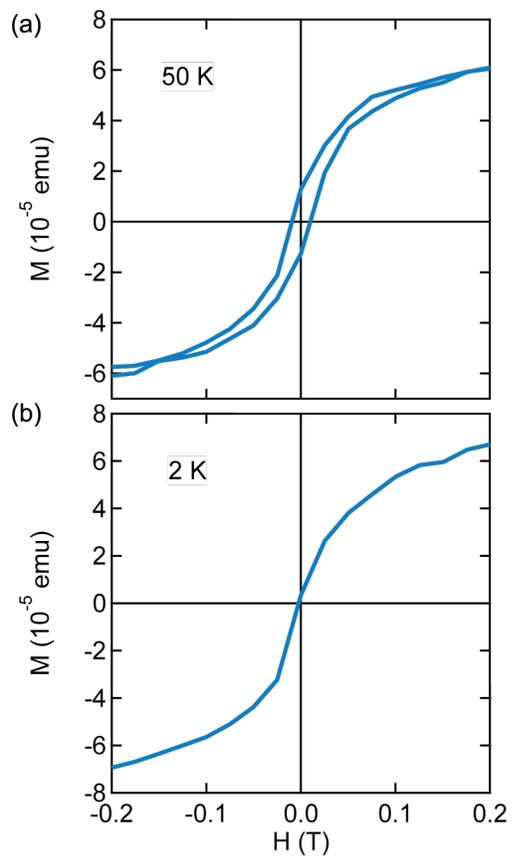

**Figure 5**

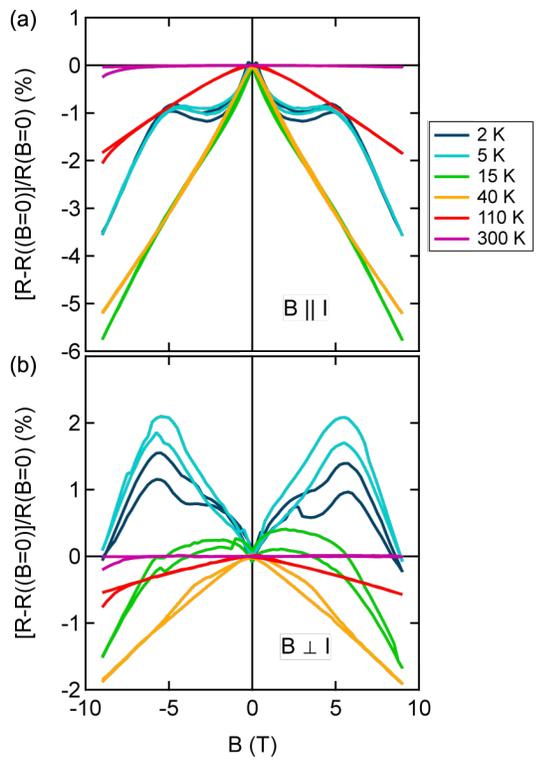

**Figure 6**

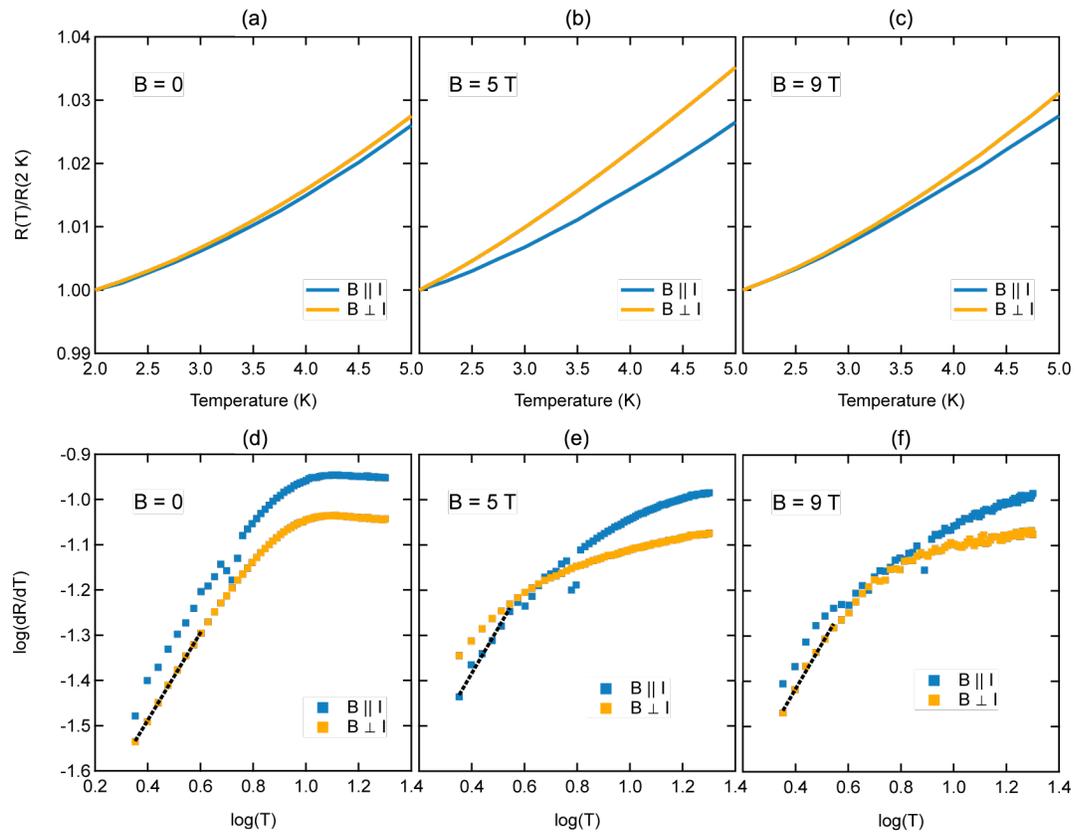

**Figure 7**

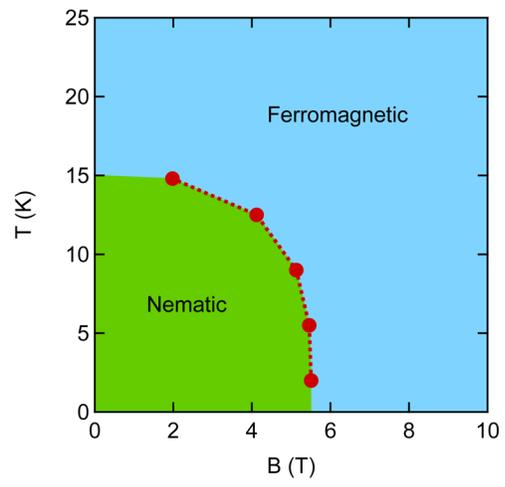